\documentclass[3p,times,procedia,dvipsnames]{elsarticle}
\usepackage{nupha_ecrc}
\usepackage[utf8]{inputenc}

\usepackage{xcolor}
\usepackage{hyperref}

\definecolor{uibred}{RGB}{167, 38, 47}

\definecolor{sbred}{RGB}{153,0,0}
\definecolor{dhbred}{RGB}{89,13,8}
\definecolor{hbred}{RGB}{198,24,38}
\definecolor{hbwhite}{RGB}{244,241,234}

\definecolor{MyDarkGreen}{rgb}{0.0, 0.30, 0.00}
\definecolor{MyDarkBlue}{rgb}{0.0, 0.00, 0.7}

\definecolor{ssred}{RGB}{224,132,116}
\definecolor{ssgreen}{RGB}{113,188,135}
\definecolor{ssblue}{RGB}{114,168,219}

\usepackage{graphicx}
\usepackage{mathtools}
\mathtoolsset{showonlyrefs} 
\usepackage{bm}
\usepackage{tikz}
\usetikzlibrary{arrows,shapes}
\usetikzlibrary{shapes.callouts,decorations.pathmorphing}
\usetikzlibrary{calc,positioning}

\def\x{{\bm x}}

\def\ubr#1#2{\underbrace{#1}_{\text{#2}}}

\newcommand{\Fig}[1]{Fig.~\ref{#1}}

\def\x{\mathbf{x}}

\newcommand{\tauekt}{\tau_\textsc{ekt}}
\newcommand{\tauhydro}{\tau_\text{hydro}}

\volume{00}

\firstpage{1}

\journalname{Nuclear Physics A}

\runauth{}

\jid{nupha}

\jnltitlelogo{Nuclear Physics A}

\usepackage{amssymb}

\usepackage[figuresright]{rotating}

\usepackage{tikz}
\newcommand{\subfig}[2]{%
\begin{tikzpicture}%
\node[rectangle] (image) at (0,0) {#2};
\node[anchor=south west] (label) at (image.south west) {(#1)};
\end{tikzpicture}%
}

\begin{document}

\begin{frontmatter}



\dochead{XXVIIth International Conference on Ultrarelativistic Nucleus-Nucleus Collisions\\ (Quark Matter 2018)}

\title{Initial conditions for nuclear collisions: theory overview}


\author{Aleksas Mazeliauskas}

\address{Institut f\"{u}r Theoretische Physik, Universit\"{a}t Heidelberg, Philosophenweg 16, D-69120 Heidelberg, Germany}

\begin{abstract}
We overview the current status and recent developments on initial conditions in ultra-relativistic nucleus-nucleus collisions. Specifically, we look at the progress in understanding the role of sub-nucleonic fluctuations in large and small collision systems. Next, we review the current ideas of going beyond boost invariant initial conditions and introducing physically motivated rapidity fluctuations at RHIC and LHC energies. Finally, we discuss the time evolution of initial stages and the matching between different descriptions of the QGP.
\end{abstract}


\end{frontmatter}


\section{Introduction}

Initial stages describe the short far-from-equilibrium phase of ultra-dense QCD matter created in heavy ion collisions.
Extensive hadronic data to model  comparisons support the existence of the hydrodynamically flowing Quark Gluon Plasma, which is formed at the end of initial stages in nucleus-nucleus collisions.
The medium properties of this new form of matter, e.g.\ transport coefficients,
are inferred from the medium response to the initial geometric fluctuations, e.g.\ via flow harmonics $v_n$.
 Therefore a detailed modeling of initial conditions in nuclear collisions is essential to the success of heavy ion physics program.
The far-from-equilibrium QCD dynamics right after the initial impact contains no less fascinating physics. The apparent fast hydrodynamization of QGP in nucleus-nucleus collisions is an actively studied topic.
Recent experimental results on multiple flow-like signals in proton-nucleus and proton-proton  collisions, put into question whether a locally equilibrated Quark Gluon Plasma could be formed even in small collision systems.

Our understanding of initial conditions is constantly evolving and, naturally, competing interpretations of the same phenomena exist. For the full range of ideas pertaining to the physics of initial conditions please refer to recent reviews \cite{Busza:2018rrf,Nagle:2018nvi,Romatschke:2017ejr,Schlichting:2016sqo} and other contributions in this issue.

\section{Fluctuations in the transverse plane}
\begin{figure}
\centering
\subfig{a}{\includegraphics[width=0.35\linewidth]{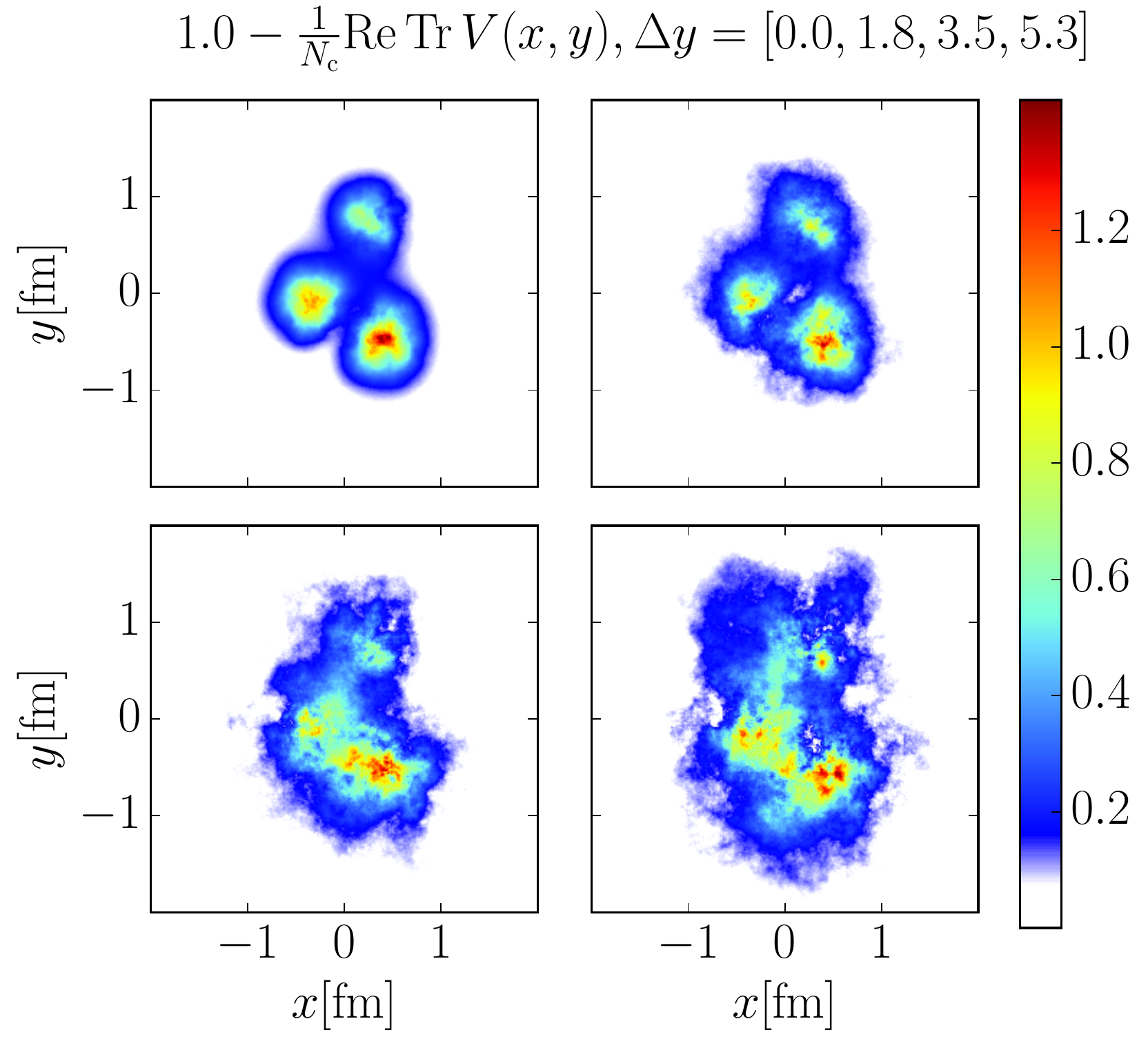}}\subfig{b}{\includegraphics[width=0.6\linewidth]{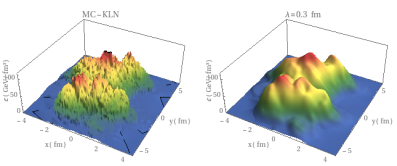}}
\caption{(a) The sub-nucleonic transverse structure of a proton. The gluonic clouds surrounding valence quarks are evolved in rapidity by JIMWLK evolution, which considerably smears and broadens the profile. Figure taken from \cite{Mantysaari:2018zdd} (b) The smearing of sub-nucleonic structure in initial density profile for non-central Pb-Pb collision at $\sqrt{s_{NN}}= 2.76\,\text{TeV}$. 
MC-KLN initial energy density profile shown before and after filtering fluctuations  of size $<0.3\,\text{fm}$, which leave most final flow observables unchanged. Figure adapted from \cite{Gardim:2017ruc}\label{fig:2d}
}
\end{figure}

The basic picture of heavy ion collisions is given by Monte-Carlo Glauber model, which describes nuclei as randomly distributed constituent nucleons, which are sampled according to the measured charge density function~\cite{Miller:2007ri}. 
Then the event-by-event fluctuations of the transverse collision geometry is a result of random positions of wounded nucleons.
Virtually all initial state models incorporate the wounded nucleon positions to determine the transverse geometry of the collision, however the physical mechanism to distribute the released collisional energy at participant positions is model dependent. 

To name just two well known examples, the IP-Glasma model uses impact parameter saturation dipole model (IP-Sat) fitted to HERA data of proton structure to model collisions between individual nucleons and to determine the averaged squared color charge density in the transverse area~\cite{Schenke:2012wb,Schenke:2012fw}. Then gaussian sampled color charges are used as sources for color-electric and color-magnetic fields, which are evolved using 2+1D classical Yang Mills equations of motion. The evolved components of gluon energy-momentum tensor are then used to initialized a hydrodynamic evolution at $\tau=0.4\,\text{fm}/c$~\cite{Ryu:2017qzn}.
In the second example, the EKRT model, the incoming nuclei produce mini-jets, whose cross-section is calculated by using colinear factorization of NLO pQCD and nuclear effects are included through parton distribution functions ~\cite{Niemi:2015qia, Niemi:2015voa}. 
The saturation condition is imposed by balancing mini-jet production with mini-jet fusion from which a local energy density is determined. Because the formation time is energy dependent, the initial density profile is evolved for a short time in a simplified 1D expansion, before passing it to a full hydrodynamic simulation at  $\tau=0.2\,\text{fm}/c$~\cite{Eskola:2017bup}.
The important feature of these initial state models is the ability to describe centrality, center-of-mass energy and collision system dependence. However the overall normalization, i.e.\ the absolute entropy production in a collision, must be tuned to the charged particle multiplicity at least in one event class~\cite{Ryu:2017qzn,Eskola:2017bup}. The recent $\sqrt{s_{NN}}=5.44\,\text{TeV}$ Xe-Xe run  showed that the integrated flow harmonics $v_n$'s  can be successful reproduced or even predicted within these models~\cite{Eskola:2017bup,Schenke:2018fci}.

There has been recent progress in studying sub-nucleonic fluctuations, which are of particular relevance in small collision systems, e.g.\ p-Pb. This involves including sub-nucleonic structure through constituent quark, wounded quark or hot spot models~\cite{Weller:2017tsr, Bozek:2017elk,Albacete:2017ajt,Mantysaari:2017cni}, alternatively one can also study fluctuations in the effective nucleon-nucleon cross-section~\cite{Alvioli:2013vk}. An independent way of constraining such sub-nucleonic fluctuations was demonstrated in \cite{Mantysaari:2017cni}. In this model the fluctuations of valence quarks (and gluon clouds surrounding them) were used to fit incoherent diffractive vector meson production in e-p collisions at HERA. The same proton size fluctuations were then used in determining the collision geometry in p-Pb system and  flow harmonics $v_2$ and $v_3$ were successfully reproduced~\cite{Mantysaari:2017cni}. More recently, the fluctuating proton model was updated to include the small Bjorken-$x$ evolution, and to revisit IP-Sat dipole parametrization (see \Fig{fig:2d}(a))~\cite{Mantysaari:2018zdd}. It will be very interesting to see if this model can simultaneously reproduce both the the transverse and longitudinal fluctuations in p-Pb collision system.
A different approach in fixing the sub-nucleonic structure is used in Bayesian analysis studies~\cite{Moreland:2018jos}. There the number of sub-nucleonic constituents, their positions and widths are free parameters, which are adjusted in simultaneous fits to p-Pb and Pb-Pb data. Although the precise number of constituent does not seem to be well constrained, several constituents with $w\sim 0.5\,\text{fm}$ width and distributed in $r\sim 1\,\text{fm}$ radius are preferred~\cite{Moreland:2018jos}.

For large collisions systems the role of sub-nucleonic fluctuations is less evident, because viscous hydrodynamic evolution smears out the small scale structure. A recent systematic investigation up to which scale $\lambda$ initial perturbation can be smeared without affecting flow observables was done in  \cite{Gardim:2017ruc} using a number of initial state models (see \Fig{fig:2d}(b)). The authors conclude that 
majority of flow observables, e.g. integrated flow harmonics $v_n$,
do not show significant dependence on small-scale structure $\lambda\leq 1\,\text{fm}$. Only  the factorization ratio $r_n(p_T^a, p_T^b)$, which measures the flow decorrelation between momentum bins, exhibit non-trivial dependence
on the smoothing length (this can be attributed to subleading flow harmonics being a response of radially excited geometry~\cite{Mazeliauskas:2015vea}). 
Hopefully, such studies will lead to a consensus, of which key features of initial conditions can be un-ambiguously constrained in nucleus-nucleus collisions.

\section{Rapidity fluctuations}

\begin{figure}
\centering
\subfig{a}{\includegraphics[width=0.5\linewidth]{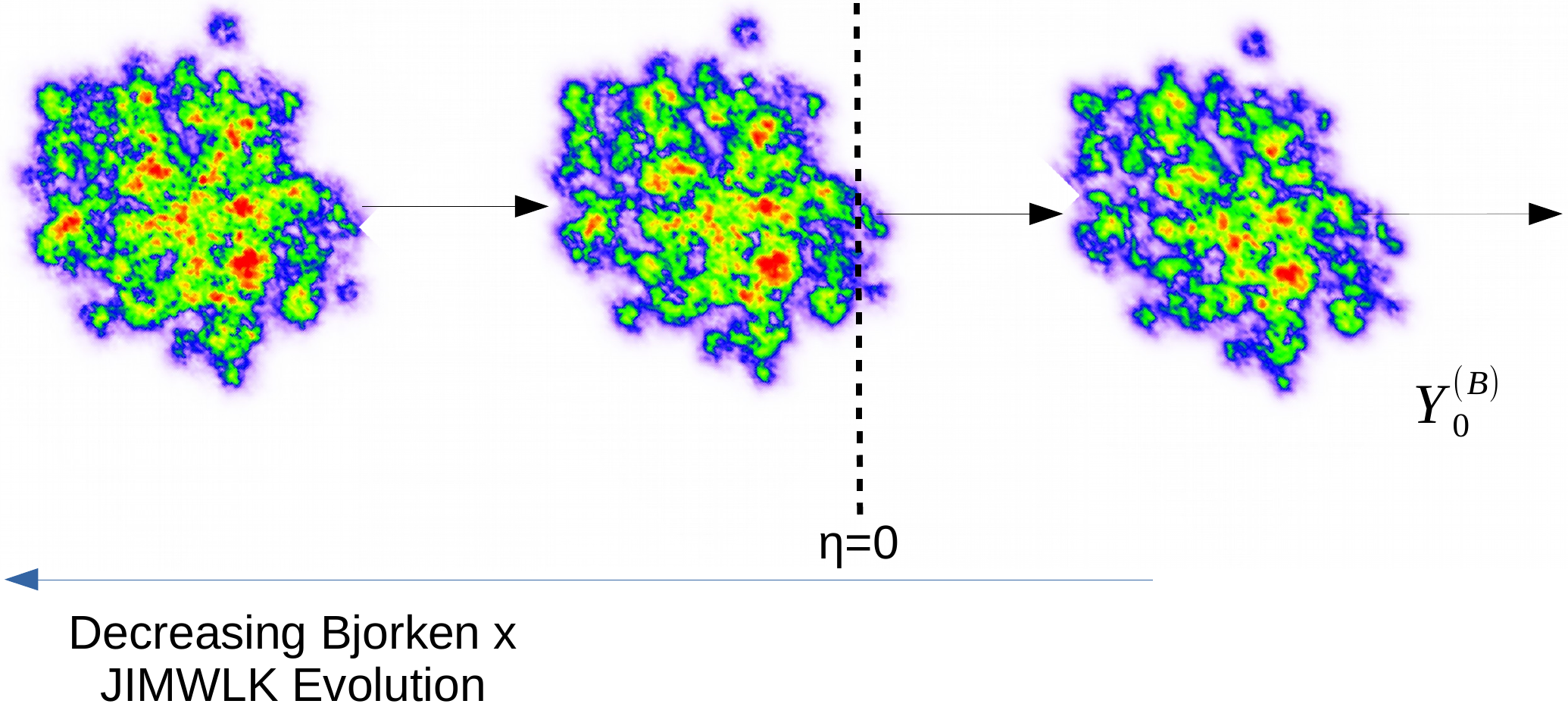}}%
\subfig{b}{\includegraphics[width=0.4\linewidth]{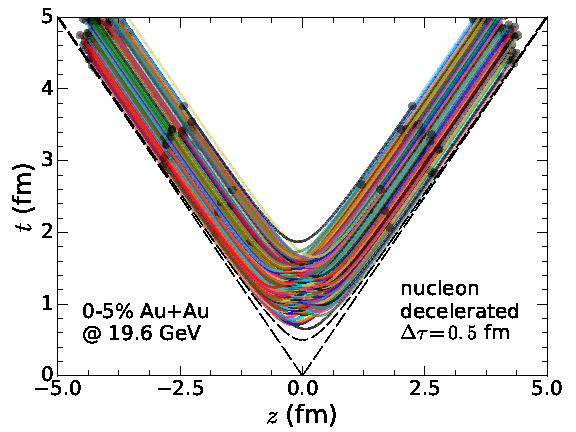}}
\caption{(a) The transverse density profile of a nucleus evolved in Bjorken $x$ by JIMWLK evolution. Figure courtesy of Scott McDonald~\cite{McDonald:2018wql}. (b) The space-time distribution of strings at their thermalization time in a Au+Au collision at $\sqrt{s_{NN}}=19.6\,\text{GeV}$.
Strings  act as external energy and flow sources for the hydrodynamic evolution, while the string ends (indicated by black dots) carry the net-baryon charge. Figure taken from~\cite{Shen:2017bsr}\label{fig:rapidity}}
\end{figure}

Ever since the observation of Bjorken that hadron production is approximately constant at mid-rapidity, boost invariant descriptions of heavy ion collisions were used in numerous theoretical studies. However, boost invariance is an emergent property of the averaged event ensemble. Even at the highest center-of-mass collision energies, there are measurable event-by-event fluctuations in the rapidity dependence of flow harmonics~\cite{Aaboud:2016jnr}. There has been a number of works explaining possible origin of such  forward-backward flow decorrelations: asymmetric energy deposition of forward and backward going wounded nucleons~\cite{Bzdak:2012tp,Bozek:2015bna}, fluctuating length of strings/flux tubes~\cite{Pang:2014pxa,Monnai:2015sca} or
quantum fluctuations of color charges~\cite{Schenke:2016ksl}. However our  understanding of physical origin of rapidity fluctuations is still much less advanced than that of the transverse geometry, and experimental results cannot be simultaneously reproduced for different settings within the same model~\cite{Bozek:2017qir}.

The first principle calculations of rapidity fluctuations at high collision energies can be done in the gluon saturation picture~\cite{Gelis:2010nm}. The highly Lorentz contracted nuclei pass each other instantaneously, filling the intermediate rapidity region by partons carrying only a fraction of the initial nucleus momentum. It was proposed in \cite{Schenke:2016ksl} to evolve the transverse profile of CGC initial conditions in rapidity using Langevin formulation of JIMWLK evolution~\cite{Lappi:2012vw}. Physically it corresponds to stochastic gluon emission with ever smaller momentum fraction (Bjorken $x$). As illustrated in \Fig{fig:rapidity}, the nuclear profile at different Bjorken $x$ values correspond to different rapidity $\eta$ with the transverse geometry increasingly smeared out. Consequently as two opposite going nuclei collide the overlap profile is no longer boost-invariant. This causes de-correlation  of initial eccentricities at different rapidity slides. This preliminary work is being followed by further studies~\cite{McDonald:2018wql}, which will hopefully establish the microscopic understanding of rapidity fluctuations with a minimal set of model parameters.  

At lower collision energies relevant to Beam Energy Scan program, the gamma factor is small and the initial collision can last for $2-3\,\text{fm}/c$, thus a dynamical initialization of the subsequent (e.g.\ hydrodynamic) medium evolution is needed. This can be done, for example, using a partonic model (e.g.\ UrQMD)~\cite{Karpenko:2015xea, Du:2018mpf} or a phenomenological string model~\cite{Shen:2017bsr}. In the latter, each binary collision of participant nucleons results in a string between the wounded nucleons. The string tension (which is a parameter of the model) slows down nucleons from the beam rapidity and convert their kinetic energy in string tension. Once a nucleon is completely stopped or after certain de-acceleration time $\Delta \tau$, string's energy, flow and baryon density (carried by the string's endpoints) is deposited in the hydrodynamic evolution as an external source (see \Fig{fig:rapidity}(b)). As a result, a range of temperatures and baryon chemical potentials are simultaneously spanned at any single time instance  in  a single collision event. Such considerations are surely very important for the extraction of the critical point location in the QCD phase diagram.

\section{Pre-equilibrium dynamics}

\begin{figure}
	\centering
\subfig{a}{\includegraphics[width=0.5\linewidth]{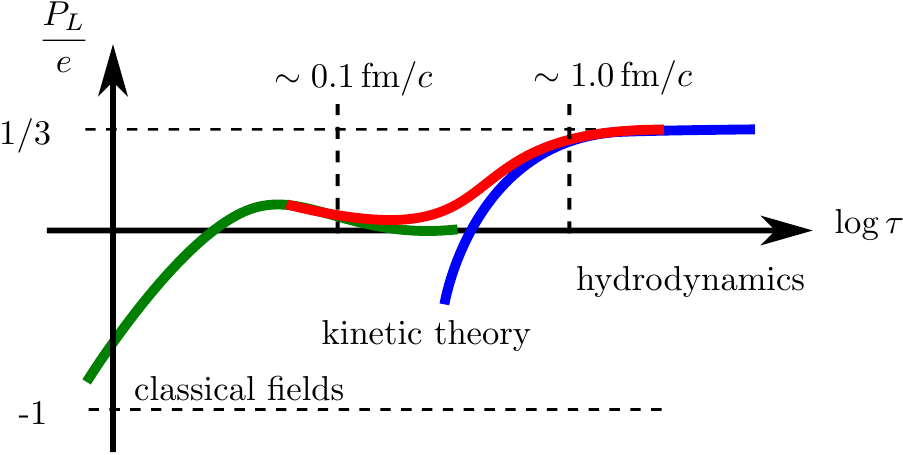}}
\subfig{b}{\includegraphics[width=0.42\linewidth]{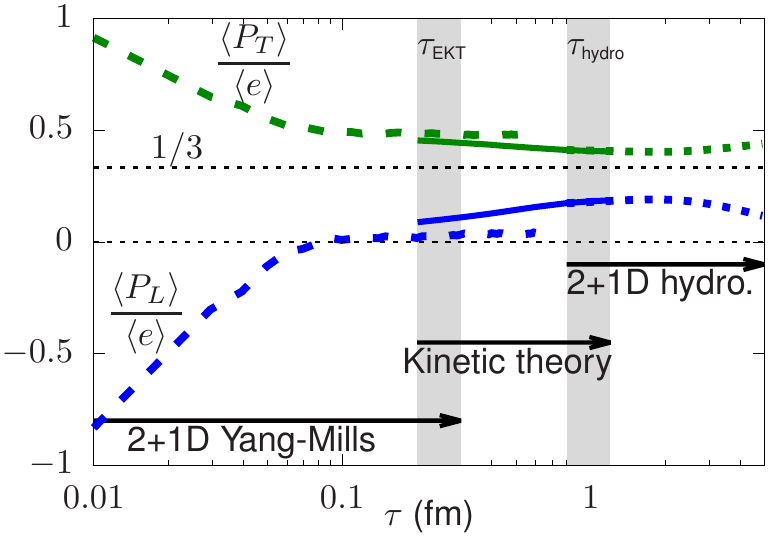}}
\caption{ (a) A proposed cartoon strategy of constructing overlapping descriptions of initial stages of heavy ion collisions by continuously matching the pressure anisotropy $P_L/e$ between classical field, kinetic theory and hydrodynamic simulations  (b) The actual realization of the matching procedure for realistic Pb-Pb initial conditions. The transversely averaged pressure is shown from a 2+1D energy-momentum tensor evolution in classical Yang-Mills, linearized kinetic theory and viscous hydrodynamics (figure taken from \cite{Kurkela:2018wud}).}
	\label{fig:earlyjf}
\end{figure}

First principle QCD calculations of the time evolution of initial conditions of heavy ion collisions is a very hard problem, which is studied only in limiting cases. Typical approaches include either using  holographic methods in the strong coupling limit applicable for certain gauge theories and, presently, not to QCD theory itself~\cite{DeWolfe:2013cua}, or effective weak coupling formulations of QCD which are strictly valid only in the asymptotic limit of very high collision energies~\cite{Gelis:2010nm,Arnold:2002zm}. 

The weak coupling scenario of the early time evolution in heavy ion collisions is  sketched in  \Fig{fig:earlyjf}(a). 
The particle production at mid-rapidity region is dominated by the interactions of small Bjorken $x$ gluons, which reach non-perturbative occupancies  $f\sim1/\alpha_s$. The resulting  strong chromo-electric and chromo-magnetic fields can be evolved in time by classical Yang-Mills equations of motion until these fields de-cohere and quantum fluctuations can be no longer neglected. In such classical field description, the longitudinal pressure component $P_L=\tau^2 T^{\eta\eta}$ starts negative, but decays towards zero, never reaching local equilibrium implied by the application of hydrodynamic evolution at later stages of the collision. In fact, because the longitudinal expansion rate for boost invariant system  diverges at early times, the hydrodynamic gradient expansion
around isotropic particle distribution function (with viscous corrections quantifying the deviations from the equilibrium) breaks down at sufficiently early times~\footnote{For the work of extending hydrodynamic descriptions to highly anisotropic systems see \cite{Alqahtani:2017mhy}.}.
 While a general theory of the equilibration process using QCD kinetic theory  has been outlined long time ago~\cite{Baier:2000sb}, up to now there were no practical frameworks to smoothly connect the early gluon production in classical field simulations with hydrodynamics of the late time plasma expansion~\cite{Keegan:2016cpi,Kurkela:2018wud,Kurkela:2018vqr}\footnote{For recent work connecting CGC initial conditions to kinetic-only description of heavy ion collisions, see~\cite{Greif:2017bnr}.}.

The QCD effective kinetic theory systematically describes the interactions between energetic quarks in gluons in Quark-Gluon Plasma~\cite{Arnold:2002zm}.
It involves $2\leftrightarrow2$ scatterings and  medium-induced  $1\leftrightarrow 2$ radiation. For a very high energy $p\gg T$ parton  transversing thermalized QGP medium, such processes describe the energy loss, i.e. jet quenching~\cite{Mehtar-Tani:2013pia}, while for in-medium  $p\sim T$ partons  they determine  the transport properties of the QGP, e.g. $\eta/s$ \cite{Ghiglieri:2018dib}. It was argued in~\cite{Baier:2000sb} that the same physical processes also bring the anisotropic CGC initial state towards equilibrium, as illustrated in \Fig{fig:earlyjf}(a).  Explicit numerical simulations show that such kinetic descriptions smoothly approaches viscous hydrodynamic behavior at late times~\cite{Kurkela:2015qoa}, while on the other hand, detailed 3+1D classical-statistical Yang-Mills simulations show the emergence of kinetic behavior at early times~\cite{Berges:2013fga}. Although the general non-equilibrium formulation of QCD kinetic theory is still lacking~\cite{Kurkela:2011ti}, valuable insight in the equilibration process can be gained from the extrapolation of current kinetic theory techniques to realistic situations~\cite{Kurkela:2015qoa}.

\begin{figure}
\centering
\subfig{a}{\includegraphics[width=0.45\linewidth]{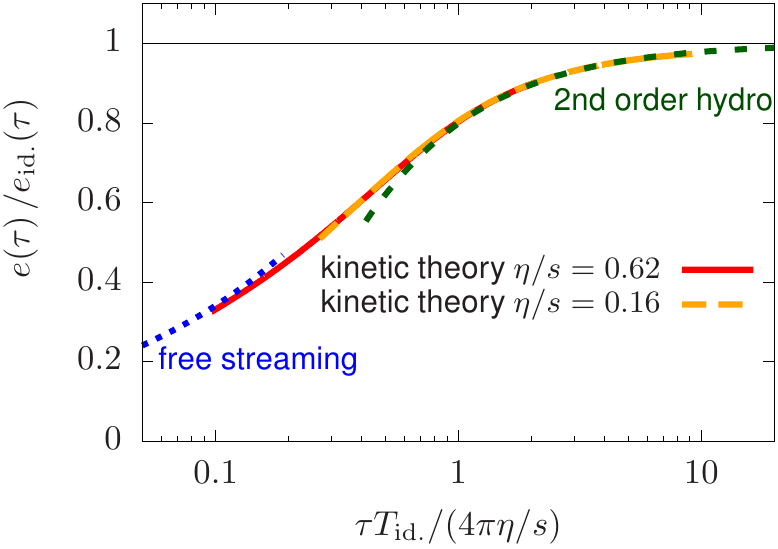}}%
\subfig{b}{\includegraphics[width=0.45\linewidth]{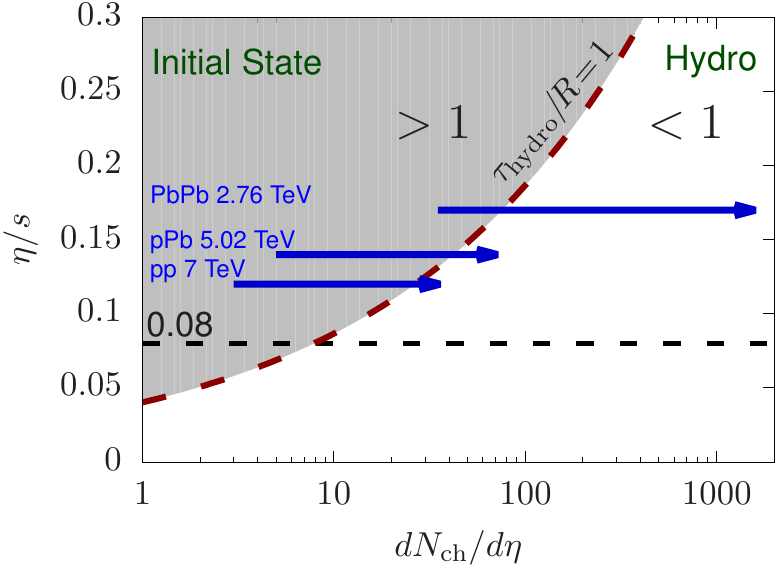}}
\caption{(a) The ratio of non-equilibrium evolution of energy density in kinetic theory and the ideal hydrodynamics for a boost invariant expansion. Kinetic theory evolution with different effective $\eta/s$ values collapse to a universal curve if written in terms of kinetic relaxation time $(\eta/s)/T$. Figure adapted from \cite{Kurkela:2018wud}  (b) The contour plot of hydrodynamization time $\tauhydro$ to system size ratio as a function of charged particle multiplicity and $\eta/s$. The white region corresponds to the cases when hydrodynamization time is smaller than the transverse system size, i.e. $\tauhydro/R<1$. Arrows indicate typical multiplicity ranges for different collision systems.\label{fig:tauR}}
\end{figure}

One of the important questions of pre-equilibrium dynamics is determining the time $\tauhydro$ when the system can be described by macroscopic hydrodynamic evolution. Shockingly, this can  happen even when the system is still rather anisotropic, i.e.  $P_L/P_T\sim 0.5$.
Actually, the hydrodynamization time for homogeneous boost invariant systems becomes independent of the initial conditions or microscopic details, if it is expressed in units of kinetic relaxation time $\tau_R\sim (\eta/s)/T$~\cite{Heller:2016rtz}.  As it can be seen from  \Fig{fig:tauR}(a) the  QCD kinetic theory evolution and 2nd order hydrodynamics starts agreeing at~\cite{Kurkela:2018wud,Kurkela:2018vqr}
\begin{equation}
\tauhydro \approx 4\pi \frac{\eta/s}{T}.
\end{equation}
One can then derive a simple formula for the hydrodynamization time ratio over the system size $\tauhydro/R$, which only depends on the charged particle multiplicity and medium constants~\cite{Kurkela:2018wud}
\begin{equation}
\frac{\tauhydro}{R}\approx\left( \frac{4\pi(\eta/s)}{2} \right)^\frac{3}{2}  \left( \frac{ dN_{\rm ch}/d\eta  }{63 } \right)^{-\frac{1}{2}}\!\!\left( \frac{S/N_\text{ch}}{7} \right) \left( \frac{\nu_{\rm eff}} {40}\right)^{\frac{1}{2}}.
\end{equation}
For $\tauhydro/R<1$ the system can reach hydrodynamic behavior while still expanding longitudinally. However, as soon as evolution time becomes comparable to the transverse system size, a much faster three dimensional expansion takes over and a long lived hydrodynamic phase could not be formed in a collision.
 As the ratio is very sensitive to the value of $\eta/s$ in the hydrodynamic phase, \Fig{fig:tauR}(b) shows a contour plot of  $\tauhydro/R$ as a function of shear viscosity and particle multiplicity (with other parameters set to their typical values). 
While in central Pb-Pb collisions $dN_\text{ch}/d\eta$ is large enough that the hydrodynamic phase can be reached for a range of $\eta/s$, small collision systems like p-Pb  and peripheral Pb-Pb are unlikely to  hydrodynamize by the time $\tau\approx R$. Indeed, one of the important signals of the formation of strongly interacting QGP phase is jet quenching, which was never observed in p-Pb collisions and, as recently clarified, is also not seen in peripheral Pb-Pb collisions~\cite{Acharya:2018njl}. 
However other signals of collective behavior are observed even in the smallest systems, which indicate that such signals of collectivity could be already formed or nearly formed in the pre-equilibirum stages. Whether it could be purely the initial state effect~\cite{Mace:2018yvl} or a result of just few rescatterings~\cite{Kurkela:2018qeb} is an ongoing debate, and the pre-equilibrium dynamics of small collisions systems will surely remain in focus.

Another very practical question in pre-equilibrium dynamics is how one 
 specifies all components of the energy-momentum tensor $T^{\mu\nu}(\tau,\x)$ needed for hydrodynamic initialization. Current approaches includes specifying only the initial energy density profile with no initial pre-flow~\cite{Niemi:2015qia, Niemi:2015voa}, using the
energy and flow from free streaming pre-evolution~\cite{Moreland:2014oya, Broniowski:2008qk, Liu:2015nwa}, passing over the energy-momentum tensor from classical field simulations~\cite{Ryu:2017qzn}, or using holography inspired initial conditions~\cite{vanderSchee:2013pia}.  
A new approach  of pre-equilibrium evolution of the transverse geometry using QCD kinetic theory was presented in~\cite{Keegan:2016cpi,Kurkela:2018wud,Kurkela:2018vqr}.
For pre-equilibrium evolution times much shorter than the transverse system size (typical situation in nucleus-nucleus collisions), initial perturbations in the energy-momentum tensor $\delta T^{\mu\nu}$ can be propagated by linear  response functions $G^{\mu\nu}_{\alpha \beta}$
\begin{equation}
\ubr{	\delta 
	T^{\mu\nu}_\x(\tauhydro,\x')}{passed over to hydro} = \int 
	d^2\x'~\ubr{G^{\mu\nu}_{\alpha 
		\beta}\left(\x-\x',\tauhydro,\tauekt\right)}{linear response function}\ubr{\delta 
	T_\x^{\alpha\beta}(\tauekt,\x')}{initial conditions}.
\end{equation}
Considering just initial energy $\delta T^{\tau \tau}$ and momentum perturbations $\delta T^{\tau i}$, this amounts to 10 independent scalar Green's functions calculable in kinetic theory~\cite{Kurkela:2018vqr}. 
Combining the non-equilibrium kinetic theory response functions with the universal background equilibration shown in \Fig{fig:earlyjf}(a), one obtains a simple practical tool to actually simulate the equilibration of far-from-equilibrium initial conditions in heavy ion collisions~\cite{kompost_github}. 
Applying this procedure to a realistic IP-Glasma initial state realizes a long sought aim~\cite{Kurkela:2016vts} of building a continuous, overlapping descriptions of initial stages in heavy ion collisions,  as  demonstrated in \Fig{fig:earlyjf}(b). The seamless matching of initial conditions to the hydrodynamic evolution could be of particular importance for the reliable  extraction of transport coefficients of the QGP~\cite{Kurkela:2018qeb}.
\section{Summary}

Initial conditions of nucleus-nucleus collisions require describing the collision dynamics in all 4 space-time coordinates immediately after the impact.
Our best understanding is of the transverse plane geometry. Hydrodynamic simulations of heavy ion collisions can successfully reproduce and even predict mid-rapidity flow harmonics across many collision systems. The studies of small systems also reveal the non-trivial structure of a proton, which can be  independently constrained by experimental data. Equally important to these efforts, is a systematic quantification of which features of initial state geometry in nucleus-nucleus collisions can be un-ambigiously constrained from the experimental observables.

Another exciting avenue of theoretical research is understanding the rapidity fluctuations in the initial conditions. At high collision energies, a significant progress was made by including first principle calculations of the rapidity evolution of the transverse profile of a collision. Hopefully,  the model to data comparisons of rapidity dependent observables will soon be on par with the studies at mid-rapidity.
At lower energy collisions, no first principle QCD description seems to be applicable and a more phenomenological approaches have to be used instead.
Because of its importance to Beam Energy Scan program, realistic modeling of initial baryon density distribution remains a high priority in heavy ion physics.

Finally, a very active line of research has been the study of early time dynamics and approach to equilibrium in large and small collision systems. This led to  deeper understanding of universality of hydrodynamic behavior away from equilibrium. Furthermore, practical tools now exist to connect the classical field description of initial state and the hydrodynamic evolution at late times in nucleus-nucleus collisions using QCD kinetic theory. There is no doubt that a comprehensive description of pre-equilibrium stages will be indispensable in explaining the observed signals of collectivity  in small collision systems.

\section*{Acknowledgments}
I would like to thank J.~Berges, P.~Bożek, K.~Eskola, S.~Floerchinger, J.~Noronha, B.~Schenke, C.~Shen, M.~Strickland, R.~Venugopalan, and U.~Wiedemann, and my collaborators A.~Kurkela, S.~Schlichting, J.-F. Paquet and D.~Teaney for discussions on these topics. I would also like to thank the Quark Matter 2018 organizers for the invitation and young researcher support.  The travel support of  HGS-HIRe made possible thanks to EMMI and QM2014 is gratefully acknowledged. 
This work was supported in part by the German Research Foundation (DFG) 
Collaborative Research Centre (SFB) 1225 (ISOQUANT).





\bibliographystyle{elsarticle-num}
\bibliography{master.bib}







\end{document}